\documentclass[preprint, 5p, twocolumn]{elsarticle}
\journal{Physics Letters A}
\usepackage[colorlinks=true, citecolor=red, urlcolor=blue ]{hyperref}
\usepackage{epsfig,newlfont,bm,subfigure,palatino,mathtools,braket,times,soul,setstack}
\usepackage{color}
\usepackage[utf8]{inputenc}
\usepackage[sc,osf]{mathpazo}
\usepackage[normalem]{ulem}
\usepackage[T1]{fontenc}
\usepackage{multirow}
\usepackage{graphics}
\usepackage{graphicx}
\usepackage{latexsym}
\usepackage{amssymb}
\usepackage{color}
\usepackage{graphics,epstopdf}
\usepackage{capt-of}
\usepackage{lipsum}
\usepackage{adjustbox}
\usepackage[table,xcdraw]{xcolor}
\usepackage{physics}
\usepackage{ragged2e}
\usepackage{comment}
\usepackage{widetext}
\usepackage[font=small,labelfont=bf,justification=justified,singlelinecheck=false]{caption}

\begin{document}
	
\title{Perfect transfer of arbitrary continuous variable states across optical waveguide lattices}	
	
\author{Tonipe Anuradha$^{1,2}$, Ayan Patra$^1$, Rivu Gupta$^{1, 3}$, Aditi Sen(De)$^1$}
	
\address{$^1$ Harish-Chandra Research Institute,  A CI of Homi Bhabha National Institute, Chhatnag Road, Jhunsi, Prayagraj - $211019$, India }
\address{$^2$ International Centre for Theory of Quantum Technologies, University of
Gda´nsk, 80-309 Gda´nsk, Poland}
\address{$^2$ Dipartimento di Fisica “Aldo Pontremoli,” Università degli Studi di Milano, I-$20133$ Milano, Italy}
	
\begin{abstract}
We demonstrate that perfect state transfer can be achieved in an optical waveguide lattice governed by a Hamiltonian with modulated nearest-neighbor couplings. In particular, we report the condition that the evolution Hamiltonian should satisfy to achieve perfect transfer of any continuous variable input state.
The states that can be transmitted need not have any specific properties - they may be pure or mixed, Gaussian or non-Gaussian in character, and comprise an arbitrary number of modes. We illustrate that the proposed protocol is scalable to two- and three-dimensional waveguide geometries. 
With the help of local phase gates on all the modes, our results can also be applied to realize a SWAP gate between mirror-symmetric modes about the center of the waveguide setup. 
\end{abstract}


\maketitle

\section{Introduction}
\label{sec:Intro}


Transferring quantum states from one portion of the system to another is essential to quantum circuits and quantum communication networks  \cite{Plenio_NJP_2004, nielsen_2010}. The discovery of quantum teleportation \cite{Bennett_PRL_1993} demonstrated that no classical protocol can be used to convey quantum information,  thereby establishing quantum advantage. Several variations of information transmission protocols have been developed including entanglement transmission \cite{Cirac_PRL_1997}, relocation of trapped ion qubits \cite{Kielpinski_Nature_2002, Seidelin_PRL_2006}, and qubit-cavity coupling \cite{Wallraff_Nature_2004, Majer_Nature_2007, Herskind_NP_2009, Paik_PRL_2011}, which however, require the in-situ manipulation of the resource \cite{Benenti_2004} and are, therefore, prone to decoherence \cite{Schoelkopf_Nature_2008}. Towards this end, the concept of quantum state transfer (QST) through dynamics was introduced  \cite{Bose_PRL_2003} to circumvent the aforementioned disadvantages. The process of state transfer comprises a substrate, such as an unmodulated spin chain with limited access, and the evolution of the system results in the transmission of an unknown qubit from one point to another \cite{Burgarth_NJP_2005, Burgarth_NJP_2005_2, Burgarth_PRA_2005, Burgarth_arXiv_2007}. The protocol allows for the reduction of errors and environmental influences since it is not necessary to apply multiple quantum gates, each of which would invariably introduce imperfections \cite{Yung_PRA_2005,Devitt_RPP_2013}. The scalability of the QST mechanism makes it a versatile tool for transferring quantum information across different nodes in a quantum circuit \cite{Raussendorf_PRL_2001, Zhou_PRL_2002, Benjamin_PRA_2004, Kay_PRA_2008, Mkrtchian_PLA_2008}. In systems with finite degrees of freedom,  QST has been demonstrated with the help of flux qubits \cite{Lyakhov_NJP_2005, Lyakhov_PRB_2006} and charge qubits \cite{Romito_PRB_2005, Tsomokos_NJP_2007} through Josephson junctions, with excitons \cite{damico_arXiv_2005, Spiller_NJP_2007} and electrons in quantum dot setups, and also using NMR techniques \cite{Zhang_PRA_2005, Zhang_PRA_2006, Fitzsimons_PRL_2007, Rao_PRA_2014}, coupled cavity qubits \cite{Paternostro_PRA_2005, Almeida_PRA_2016},  superconducting qubits \cite{Majer_Nature_2007, Sillanpaa_Nature_2007, Togan_Nature_2010, You_Nature_2011}, and  trapped ions \cite{Schmidt-Kaler_Nature_2003, Leibfried_Nature_2003}.

Since the primary goal of QST is to transmit an unknown quantum state, it is essential that the output state be as close to the input state as possible. The success of the protocol is thus determined by the  fidelity \cite{Uhlmann_RMP_1976, Jozsa_JMO_1994} between the initial and final states, with a higher value indicating a more faithful and reliable transfer of states. A natural question that arises, is whether such systems can be engineered, which can ensure that the output state is exactly the same as the input state, known as perfect state transfer (PST). In the case of an array of interacting spins, it was shown that the couplings between the spins when doctored to specific values \cite{Albanese_PRL_2004, Christandl_PRA_2005, Kay_PRA_2006, DiFranco_PRL_2008, Nguyen_ANS_2010,Nikolopoulos_JPCM_2004, Karbach_PRA_2005, Ying_CTP_2007, Gualdi_PRA_2008, Gualdi_NJP_2009, Shi_PRA_2015} can lead to PST. In recent years, several approaches for accomplishing PST have been made  which include adjusting couplings in the boundaries \cite{Banchi_PRL_2011,Banchi_PRA_2010,Wojcik_PRA_2005,Korzekwa_PRA_2014,Zwick_NJP_2014,Agundez_PRA_2017} as well as spins \cite{Pemberton_PRL_2011,Karimipour_PRA_2012,Bayat_PRA_2014}, dual(multi)-rail encoding \cite{Burgarth_PRA_2005,Burgarth_NJP_2005,Shizume_PRA_2007}, and manipulations of external magnetic fields \cite{Fitzsimons_PRL_2006,Eckert_NJP_2007,Murphy_PRA_2010,Lorenzo_PRA_2013,Zhang_PRA_2005}, to mention a few.

Beyond systems with finite degrees of freedom, continuous variable (CV) systems \cite{Adesso_OSID_2014, Serafini_2017}, characterized by canonically conjugate quadrature variables possessing an infinite spectrum, have emerged as potential candidates for the realization of several information theoretic protocols such as dense coding \cite{BraunsteinKimbleDC}, teleportation \cite{Braunstein_PRL_2000}, quantum cloning \cite{andersen2005}, preparation of cluster states for one-way quantum computation \cite{Yoshikawa_APL_2016} etc. Within this framework, optical lattices realized through coupled optical waveguides in a one-dimensional array, have been employed to revolutionize the implementation of quantum computation \cite{Slussarenko_APR_2019} and the transfer of non-classical features \cite{Rai_PRA_2008, Longhi_PRL_2008}. 
 They constitute an experimentally feasible mechanism for manipulating  light \cite{Takesue_Optica_2008, Camacho_Optica_2012, Das_PRL_2017, Kannan_SA_2020, Zhang_Nature_2021} or  simulating quantum spin models using optical systems \cite{Hunh_PNAS_2016, Bello_PRX_2022}. Such waveguide arrays are fabricated using femtosecond laser  techniques \cite{pertsch_2004_discrete,itoh_2006_ultrafast, Szameit_OE_2007, Szameit_JPB_2010, meany_2015} and nanofabrication methods \cite{Rafizadeh_CLE_1997, Belarouci_JL_2001},  and they have high resistance to noise \cite{Perets_PRL_2008, Dreeben_QST_2018}, thereby preserving coherence and polarization \cite{Gattass_NP_2008, Valle_JOA_2009, Smith_OE_2009, Sansoni_PRL_2010, Lepert_OE_2011} up to a high degree of accuracy.
Although the transfer of both qubits and qudits \cite{Latmiral_PRA_2015} as well as  pure CV states \cite{Rodriguez-Lara_JOSAB_2014} has already been studied in this system, the perfect transmission of arbitrary CV states by manipulating interaction between waveguide arrays has not been reported yet. Moreover, experimental success has been limited to the transmission of single-photon Fock and path-entangled states \cite{Bellec_OL_2012, Perez-Leija_PRA_2013, Chapman_Nature_2016},  two-mode squeezed vacuum state \cite{Swain_JO_2021}, and two-photon N$00$N states \cite{rai_2010_quantum}. \textcolor{black}{The transfer of quantum properties inherent in the two most resourceful CV states - the coherent and the squeezed states have also been demonstrated using circularly coupled waveguide arrays in four-mode setups~\cite{Rai_JO_2022}.} 

In this paper, we present a waveguide Hamiltonian composed of nearest-neighbor (NN) interactions, whose evolution allows for the perfect transfer of any arbitrary multimode CV state, pure or mixed, from one end of the lattice to the other.
Mirror symmetry has been demonstrated to be useful in achieving and explaining PST in both discrete \cite{Christandl_PRA_2005, Qian_PRA_2005} as well as CV systems \cite{Chapman_Nature_2016, Bellec_OL_2012}. In a similar spirit, we provide the necessary conditions that the evolution via mirror-symmetric waveguide Hamiltonian must satisfy in order to facilitate PST.  More specifically, given a waveguide lattice consisting of a fixed number of modes, we specify the strengths of the NN couplings which must be engineered (prior to the PST protocol) so that the system supports the PST procedure. Our protocol is scalable to two- as well as three-dimensional lattices, where PST is possible to modes placed mirror symmetrically opposite to the input modes, about the center of the lattice. \textcolor{black}{We explicitly provide the phase gates to be applied at the output mode to accomplish PST in arbitrary lattices. Importantly, the gates are state-independent and only depend on the structure of the waveguide setup.}  We also indicate the possible realization of the coupling coefficients in waveguide circuits. We establish a connection between the state transfer protocol and the generation of entanglement between the input and the output modes, which might shed some light on the hitherto unexplained phenomenon of PST. With the help of additional local phase shifts, we illustrate that the Hamiltonian can be used to design a SWAP gate between modes equidistant from the center of the waveguide arrangement. 


Our paper is organized in the following way. We begin with a brief recapitulation of the state transfer protocol for CV systems in Sec. \ref{sec:qst_cv}. In this section, we also derive the condition that must be satisfied by the evolution Hamiltonian to support PST. Sec. \ref{sec:our_waveguide} deals with the PST protocol for one-dimensional waveguide arrays and connects the PST protocol with the mode-entanglement. In Sec. \ref{sec:sec_swap}, we describe how to use the Hamiltonian in order to implement a SWAP gate between mirror-symmetric modes. The realization of the PST protocol through waveguide arrangements in higher dimensions is presented in Sec. \ref{sec:2d-PST} and we further discuss the possible experimental realization of the nearest-neighbor modulated couplings. We end our paper with conclusions and discussions in Sec. \ref{sec:conclu}.

\section{Framework for State transfer in continuous variable systems}
\label{sec:qst_cv}
Within the framework of continuous variables, quantum state transfer involves the transport of an arbitrary or a fixed CV state over a quantum network of modes. We deal with a quantum network consisting of a finite number of modes that are connected only through nearest-neighbor (NN) couplings, in different spatial dimensions. All modes except the first one are initially in the vacuum state $|0\rangle$, and an arbitrary state $\rho_{\text{in}}$ is localized in the first mode. One of the main features of the protocol is that access to the input and output modes is enough to guarantee state transfer. Note that even multimode states can undergo transfer across the network, in which case the number of accessible input and output modes must be more. In the situation under consideration, the $N$-mode initial state takes the form as
\begin{equation}
    \rho(0) = \rho_{\text{in}} \otimes |0\rangle \bra{0}^{\otimes N - 1}.
    \label{eq:initial_state}
\end{equation}
Let the Hamiltonian which governs the evolution of the system be $\hat{H}$. Under the effect of  $\hat{U} = \exp (- i \hat{H} t)$, the time evolved state, at time $t$, becomes $\rho(t) = \hat{U}~ \rho (0)~ \hat{U}^\dagger$. Since the principal goal is to transfer the input state to the desired output mode as accurately as possible, the figure of merit for the process is the fidelity between the initial and final states, i.e., to check how close the states $\rho_{\text{in}}$ and $\rho_N (t) = \Tr_{1, 2, \dots, N - 1} \rho(t)$ are to each other. The quantum fidelity between two states $\rho_1$ and $\rho_2$ is given by the Ulhmann fidelity \cite{Uhlmann_RMP_1976},
\begin{equation}
    \mathcal{F}(\rho_1, \rho_2) = \Big( \Tr \sqrt{\sqrt{\rho_1} \rho_2 \sqrt{\rho_1}} \Big)^2,
    \label{eq:Ulhmann_fid}
\end{equation}
where we identify $\rho_{\text{in}}$ and $\rho_N(t)$ as $\rho_1$ and $\rho_2$ respectively. For two generic single-mode Gaussian states, i.e., displaced squeezed thermal states, Eq. \eqref{eq:Ulhmann_fid} has been computed exactly \cite{Mandarino_IJQI_2014} and, in the phase-space picture, has the form
\begin{eqnarray}
    \mathcal{F}(\rho_1, \rho_2) = \frac{\exp [-(\mathbf{d}_1 - \mathbf{d}_2)^T (\Xi_1 + \Xi_2)^{-1} (\mathbf{d}_1 - \mathbf{d}_2) ]}{\sqrt{\Delta + \delta} - \sqrt{\delta}},
    \label{eq:fid_exp}
\end{eqnarray}
where $\mathbf{d}_{j}$ and $\Xi_{j}$ are the displacement vector and covariance matrix  corresponding to the single-mode Gaussian state $\rho_j$ with $\Delta = \det(\Xi_1 + \Xi_2)$ and $\delta = 4 \prod_{j = 1}^2 (\det(\Xi_j) - 1/4)$ respectively. We elucidate the phase-space formalism for CV Gaussian states in Appendix \ref{app:CV}. The fidelity in Eq. \eqref{eq:fid_exp} is a function of time since one of the involved states is the time-evolved state in mode $N$.

Perfect state transfer (PST) is defined by maximizing the fidelity with respect to the evolution time for a given Hamiltonian $\hat{H}$, such that at an optimal time, $t_{opt}$,  $\mathcal{F}(\rho_{\text{in}}, \rho_N (t_{opt})) = 1$. It has been shown for spin systems, that any generic unmodulated Hamiltonian can never lead to PST \cite{Christandl_PRL_2004}, and even for CV systems, Hamiltonians have been identified which allow PST  only for a restricted class of states \cite{Rai_JO_2022}. Our goal now is to determine the conditions which must be satisfied by the Hamiltonian, such that it can lead to PST of any arbitrary CV state at some $t_{opt}$.
 
\subsection{Criterion for  perfect state transfer}
\label{subsec:condition_PST}


To achieve PST, the state in the mode $N$ must become $\rho_{N} (t)  = \rho_{\text{in}}$ at some optimal time, $t_{\text{opt}}$, upto which the evolution is allowed to take place. In other words, at $t = t_{\text{opt}}$,  the final state takes the form as
\begin{equation}
    \rho (t_{opt}) = \tilde{\rho}_{1, 2, \cdots, N - 1} \otimes \rho_{\text{in}},
    \label{eq:final_state_PST}
\end{equation}
where $\tilde{\rho}_{1, 2, \cdots, N - 1}$ is the reduced state in the remaining modes. \textcolor{black}{Thus, a sufficient condition for PST to take place is that the final state satisfies Eq. \eqref{eq:final_state_PST}, i.e., it is separable in the partition between the output and the rest of the remaining modes, $1 2 \cdots N - 1:N$ bipartition. Any residual entanglement between the output state at the last mode and the remaining lattice modes may imply that $\rho_N(t) \neq \rho_{\text{in}}$ (especially if $\rho_{\text{in}}$ is pure), since the reduced subsystem of any entangled state may not be equal to that of an initial product state.}

In order to ensure that the output state is of the form given in Eq. \eqref{eq:final_state_PST}, the Hamiltonian through which the system is evolved must have a specific structure. For this purpose, we concentrate on the waveguide Hamiltonian in which manipulation and processing of light is possible. Let us consider waveguides comprising  $\mathcal{L}, \mathcal{B}$, and $\mathcal{H}$ number of modes in three dimensions, having only nearest-neighbor (NN) couplings. The corresponding Hamiltonian dictating the dynamics reads as
\begin{eqnarray}
   \nonumber \hat{H} =  && \sum_{j, k, l = 1}^{\mathcal{L} - 1, \mathcal{B}, \mathcal{H}} J^\mathcal{L}_{j} (\hat{a}_{(j,k,l)}^\dagger \hat{a}_{(j+1,k,l)} + \hat{a}_{(j+1,k,l)}^\dagger \hat{a}_{(j,k,l)}) + \\
    && \nonumber \sum_{j, k, l = 1}^{\mathcal{L}, \mathcal{B}-1, \mathcal{H}} J^\mathcal{B}_{k} (\hat{a}_{(j,k,l)}^\dagger \hat{a}_{(j,k+1,l)} + \hat{a}_{(j,k+1,l)}^\dagger \hat{a}_{(j,k,l)}) + \\
    && \sum_{j, k, l = 1}^{\mathcal{L}, \mathcal{B}, \mathcal{H}-1} J^\mathcal{H}_{l} (\hat{a}_{(j,k,l)}^\dagger \hat{a}_{(j,k,l+1)} + \hat{a}_{(j,k,l+1)}^\dagger \hat{a}_{(j,k,l)}),
    \label{eq:H_NN}
\end{eqnarray}
where we have set $\hbar = 1$ and $J^{\mathcal{L}}_{j}$ is the real coupling strength between adjacent modes, $(j,k,l)$ and $(j+1,k,l)$, along one dimension and a similar convention holds for $J^\mathcal{B}_{k}$ and $J^\mathcal{H}_{l}$ in the other two dimensions. $\hat{a}_{(j,k,l)}$ denotes the bosonic annihilation operator of the mode at site $(j,k,l)$. The total number of modes in the lattice with an open boundary condition is given by $N = \mathcal{L} \times \mathcal{B} \times \mathcal{H}$.

\subsubsection{Condition on the dynamics}
\label{subsubsec:AN1_condition}

We will analyze the evolution of the state in the Fock space notation. Since any single mode state can be represented as $\rho_{\text{in}} = \sum_{n, m = 0}^{\infty} C_{m, n} \ket{n} \bra{m}$, with $\sum_{n = 0}^\infty C_{n, n} = 1$ for normalization, we can write the initial state from Eq. \eqref{eq:initial_state} at time $t = 0$ as
\begin{equation}
   \rho (0) = \sum_{n, m = 0}^{\infty} C_{n, m} \frac{\hat{a}_{(1, 1, 1)}^{\dagger^n}(0)}{\sqrt{n!}} \Big(\ket{0}\bra{0}\Big)^{\otimes N} \frac{\hat{a}_{(1, 1, 1)}^{m}(0)}{\sqrt{m!}},
    \label{eq:in_state_fock}
\end{equation}
where $\ket{k}$ is the number state containing $k$ photons, $\hat{a}_{(1, 1, 1)}^{\dagger}$ is the creation operator for the mode at site $(1, 1, 1)$ in which the input state is provided. The time-evolved state $\rho(t)$ can be written as
\begin{equation}
    \rho(t) = \sum_{n = 0}^{\infty} C_{n, m} \frac{\hat{a}_{(1, 1, 1)}^{\dagger^n}(t)}{\sqrt{n!}} \Big(\ket{0}\bra{0}\Big)^{\otimes N} \frac{\hat{a}_{(1, 1, 1)}^{m}(t)}{\sqrt{m!}}.
    \label{eq:psi_t}
\end{equation}
Let us now find out the explicit form of $\hat{a}_{(1, 1, 1)}^{\dagger^n}(t)$. In the Heisenberg picture, the time evolution of the annihilation operator for a single mode, at site $(u, v, w)$, is given by
\begin{eqnarray}
   \nonumber \frac{d \hat{a}_{(u,v,w)}^\dagger}{dt}  = &&  -i [\hat{H}, \hat{a}_{(u,v,w)}^\dagger] \\
   \nonumber   = && - i\Big( J^\mathcal{L}_{u-1} \hat{a}_{(u-1,v,w)}^\dagger +  J^\mathcal{L}_{u} \hat{a}_{(u+1,v,w)}^\dagger \Big) + \\
   \nonumber &&  - i \Big( J^\mathcal{B}_{v-1} \hat{a}_{(u,v-1,w)}^\dagger +  J^\mathcal{B}_{v} \hat{a}_{(u,v+1,w)}^\dagger\Big) + \\
    &&  - i \Big( J^\mathcal{H}_{w-1} \hat{a}_{(u,v,w-1)}^\dagger +  J^\mathcal{H}_{w} \hat{a}_{(u,v,w+1)}^\dagger\Big),
    \label{eq:Heisenberg_ak}
\end{eqnarray}
with $i = \sqrt{-1}$. Let us consider the set of operators $\hat{\mathbf{a}}^\dagger (t) = \{\hat{a}_{(1,1,1)}^\dagger (t), \hat{a}_{(1,1,2)}^\dagger (t), \cdots \hat{a}_{(\mathcal{L},\mathcal{B},\mathcal{H})}^\dagger (t)\}^T$. The time evolution of the creation operators for the $N$ modes can be succinctly represented as
\begin{eqnarray}
    \nonumber && \frac{d \hat{\mathbf{a}}^\dagger}{dt} = - i \mathcal{M} \hat{\mathbf{a}}^\dagger(0), \\
    \implies &&  \frac{d \hat{a}_{\{q\}}^\dagger}{dt} = \sum_{\{q'\} = (1,1,1)}^{(\mathcal{L},\mathcal{B},\mathcal{H})} -i \mathcal{M}_{\{q\},\{q'\}} \hat{a}_{\{q'\}}^\dagger(0),
    \label{eq:Heisenberg_M}
\end{eqnarray}
where $\{q\}$ represents the collection of the three mode indices, i.e., $\{q\} = (u,v,w)$, \textcolor{black}{in three dimensions}, and $\mathcal{M}$ is an $N \times N$ matrix characterizing the operator evolution, which is defined as $\mathcal{M}_{\{q\}\{q'\}} = J_{\{q-1\}} \delta_{\{q'\},\{q-1\}} + J_{\{q\}} \delta_{\{q'\}, \{q+1\}}$, with $\delta$ representing the Kronecker delta function. \textcolor{black}{Here, ${q\pm1}$ denotes the addition or subtraction of one mode from any dimension. Specifically, ${q\pm1}= (u\pm1, v, w)$ or a similar adjustment in any one of the mode indices, $u, v, ~\text{or}~ w$, within the waveguide arrangement, indicating a shift from one lattice point to its nearest neighbor.} Therefore, the time-evolved annihilation operators take the form,
\begin{eqnarray}
   \nonumber && \hat{\mathbf{a}}^\dagger (t) = \exp (-i\mathcal{M} t) \hat{\mathbf{a}}^\dagger(0) = \mathcal{A} \hat{\mathbf{a}}^\dagger(0), \\
    \implies && \hat{a}_{\{q\}}^\dagger (t) = \sum_{\{q'\} = (1,1,1)}^{(\mathcal{L},\mathcal{B},\mathcal{H})} \mathcal{A}_{\{q\},\{q'\}} \hat{a}_{\{q'\}}^\dagger(0),
\end{eqnarray}
with $\mathcal{A} = \exp (-i \mathcal{M} t)$ being the evolution matrix and the evolution coefficients are given by its elements $\mathcal{A}_{\{q\},\{q'\}}$, which are functions of the system parameters $J, N$ and also of the time $t$. Using the aforementioned formalism, we can rewrite the final output state in Eq. \eqref{eq:psi_t} as
\begin{eqnarray}
   \nonumber &&  \rho(t)  = \sum_{n, m = 0}^{\infty} C_{n, m}  \frac{(\sum_{\{q'\} = (1,1,1)}^{(\mathcal{L},\mathcal{B},\mathcal{H})} \mathcal{A}_{(1,1,1),\{q'\}} \hat{a}_{\{q'\}}^\dagger(0))^n}{\sqrt{n!}} \times \\
  &&  \Big(\ket{0}\bra{0}\Big)^{\otimes N} \frac{(\sum_{\{q'\} = (1,1,1)}^{(\mathcal{L},\mathcal{B},\mathcal{H})} \mathcal{A}^{*}_{(1,1,1),\{q'\}} \hat{a}_{\{q'\}}(0))^m}{\sqrt{m!}}. ~~
     \label{eq:Heisenberg_psi}
\end{eqnarray}
In order that the state at mode $\{N\} = (\mathcal{L},\mathcal{B}, \mathcal{H})$ reduces to $\rho_{\text{in}}$ at the optimal time, we must have 
\begin{equation}
    |\mathcal{A}_{(1,1,1),\{q'\}}| = \begin{cases}
        1, & \text{when}~~ \{q'\} = \{N\} \\
        0, & \text{otherwise},
    \end{cases}
    \label{eq:A_N1_condition}
\end{equation}
at $t = t_{\text{opt}}$. Note that the vanishing of the evolution coefficients $\mathcal{A}_{(1,1,1),\{q\}}$ (for $\{q\} \neq \{N\}$) is dictated by the photon-number conserving nature of the Hamiltonian. As a result, at $t = t_{\text{opt}}$ the output state results in PST as 
\begin{equation}
   \rho(t_{\text{opt}}) = \sum_{n, m = 0}^{\infty} C_{n, m} e^{i \phi_{nm}} \frac{\hat{a}_{\{N\}}^{\dagger^n}}{\sqrt{n!}} \Big(\ket{0}\bra{0}\Big)^{\otimes N} \frac{\hat{a}_{\{N\}}^{m}}{\sqrt{m!}},
\end{equation}
where $\phi_{nm}$ is a phase accumulated during the evolution, which can be dealt with by applying appropriate phase gates at the output port. Note that, since the input and output ports are kept accessible during the PST protocol, the implementation of gates at those ports is allowed \cite{Christandl_PRA_2005}. Therefore, a waveguide Hamiltonian whose structure can satisfy the conditions specified by Eq. \eqref{eq:A_N1_condition} on the evolution coefficients, can accurately transfer any arbitrary input state from one part of the waveguide lattice to the other.

\textbf{Criterion for PST.} When an arbitrary single-mode state impinged at mode $\{q\} = (u,v,w)$ has to be \textit{perfectly} transmitted to the mode $\{q'\} = (u',v',w')$, the evolution coefficient governed by the Hamiltonian in Eq. \eqref{eq:H_NN}, must satisfy $|\mathcal{A}_{\{q\},\{q'\}}| = 1$ and $|\mathcal{A}_{\{q\},\{q''\}}| = 0$ for $\{q''\} \neq \{q'\}$, at the optimal time $t_{\text{opt}}$.

\begin{figure}
    \includegraphics[width=\linewidth]{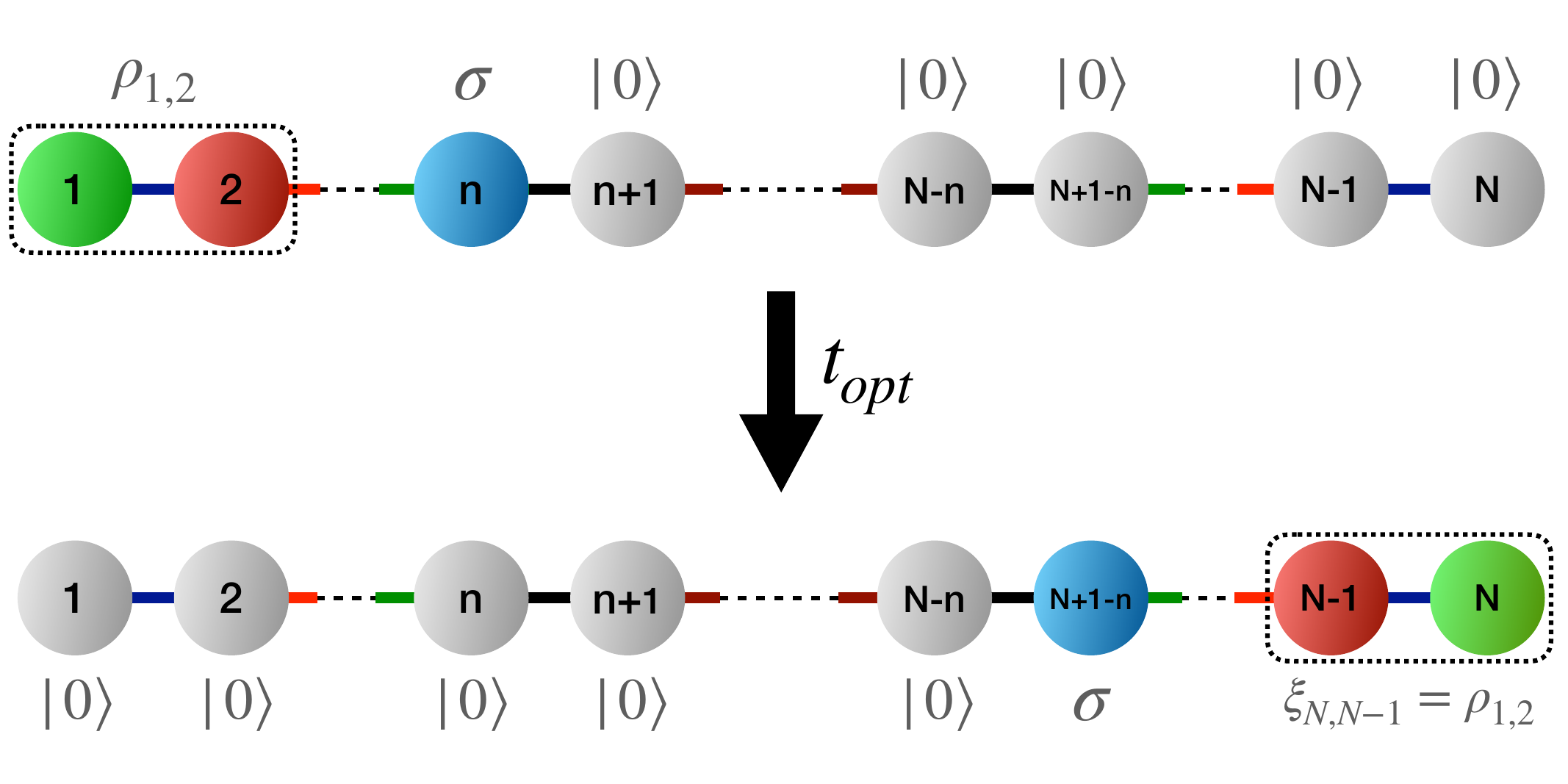}
\captionsetup{justification=Justified,singlelinecheck=false}
     \caption{\textbf{Perfect state transfer in $1D$ linear waveguide array}. The system consists of $N$ modes interacting through modulated mirror-symmetric nearest-neighbor couplings.  Any single-mode input state, $\sigma$ at mode $n$ as well as a multimode state, $\rho_{1,2}$ at modes $1$ and $2$ can be perfectly transferred, after evolution according to the Hamiltonian in Eq. \eqref{eq:H_NN} for $t = t_{\text{opt}}$, to the mode mirror-symmetric about the center of the arrangement (mode $N - n + 1$ and modes $N-1, N$ respectively) as shown. PST requires the application of appropriate local phase gates at the receivers' end, i.e., on the modes $N-n+1$, $N-1$, and $N$. Note that the intermodal non-input states that we consider to be in vacuum states can be initialized to any arbitrary CV state.}
    \label{fig:sch_pst_1D}
\end{figure}

\section{One-dimensional waveguide array  with mirror symmetry}
\label{sec:our_waveguide}
It is well established that a lattice with unmodulated nearest-neighbor couplings cannot lead to perfect state transfer \cite{Christandl_PRL_2004, Christandl_PRA_2005,coutinho_arXiv_2023} unless it consists of only $2$ or $3$ nodes. Our goal now is to find a protocol that can allow us to transport an arbitrary CV state from one end of the waveguide, with $N \geq 3$ modes, to the other end with unit fidelity. Although PST has been demonstrated in spin systems for arbitrary qubit states, a similar analysis is missing in the continuous variable regime. Previous works have identified interactions that can perfectly transfer only a restricted class of single-mode states, e.g., path-entangled states,  single photon states \cite{Perez-Leija_PRA_2013}, and two-photon N$00$N states \cite{Swain_JO_2021}. \textcolor{black}{Note, however, that the techniques developed in Ref.~\cite{PhysRevA.103.0437} can be adapted to study the perfect transfer of arbitrary CV target states in one-dimensional lattices.} In this section, we construct a Hamiltonian with modulated nearest-neighbor couplings which makes the perfect transfer of any CV state possible across the length of a one-dimensional waveguide system.

We consider a model comprising $N$ identical waveguides arranged in an array and coupled to each other (see Fig. \ref{fig:sch_pst_1D}, with coupling strengths denoted by $J_j$ between adjacent modes $j$ and $j + 1$). The coupling constants exhibit mirror symmetry between waveguide modes, i.e. $J_j = J_{N-j}$. The Hamiltonian that governs the $N$-mode system is given in Eq. \eqref{eq:H_NN}, with $\mathcal{B} = N$ and $\mathcal{L} = \mathcal{H} = 1$, where we represent $N = 2m$ for even modes and $N = 2m+1$ for odd modes, with $m$ being an integer taking values $1, 2, 3, \cdots$. We consider the coupling strength as

    \begin{eqnarray}
   J_{j}  = J \sqrt{\frac{j(N-j)}{N-1}},
   \label{eq:opt_J}
\end{eqnarray}
with $j = 0,1,2,...,m-1$ and $J$ being a constant, the value of which determines the optimal evolution time across the waveguide setup that allows for PST. We note that the coupling strength is motivated by the treatment given in Ref. \cite{Christandl_PRL_2004}, but with a normalization factor of $\sqrt{N-1}$. The additional factor ensures that the couplings do not assume very large values as the number of waveguide modes increases, which, in turn, facilitates their experimental implementation, as we will discuss later. \textcolor{black}{Note that the SU$(2)$ symmetry of two-level systems in the Schrodinger picture with the Heisenberg dynamics of the creation and annihilation operators implies that the discrete variable model~\cite{Christandl_PRL_2004} can be adapted to explain PST in CV systems~\cite{PhysRevA.103.0437, PhysRevA.105.053710}. Our discussion serves as an explicit derivation to show that this is indeed the case. Moreover, we demonstrate that the successful implementation of PST requires the application of local phase gates to the output modes. These phase corrections are independent of the specific input states being transferred and depend solely on the total number of modes in the waveguide array. Additionally, we provide the precise values of the local phases necessary to complete the PST process.}

Let us take an arbitrary $\mathcal{P}$-mode state as the input, with $\mathcal{P} \leq \lceil N/2 \rceil$, where $\lceil N/2 \rceil$ is the smallest integer greater than or equal to $N/2$. \textcolor{black}{Here, we consider the expression of any $\mathcal{P}$-mode state in the Fock basis, and not merely a Fock state.} The final state of the system, upon evolution, is characterized by Eq. \eqref{eq:Heisenberg_psi}. Since our waveguide lattice satisfies mirror symmetry about its center, it can support PST from any mode $j$ to the mode $N-j+1$. To see this, let us consider the expression for $\mathcal{A}_{j,N-j+1}$ which turns out to be
\begin{eqnarray}
   \nonumber \mathcal{A}_{j, N-j+1}&&  =  \frac{(-i)^{N-1}}{2^{j}} \sin^{N - 2j + 1} \Big( \frac{J t}{\sqrt{N - 1}} \Big) \times \\
 && \Big(2^j + \sum_{r = 1}^{j-1} (-2)^{r} c^{j,N}_r \cos^{2r} \frac{J t}{\sqrt{N - 1}}\Big),
  \label{eq:AN1_PST-1}
\end{eqnarray}
where $j = 1, 2, \cdots, \lceil N/2 \rceil$ and $c^{j,N}_r$ are integers. For the exemplary cases of $j = 1, 2,$ and $3$, the evolution coefficients have the form
\begin{eqnarray}
    \mathcal{A}_{1,N} &=& (-i)^{N-1} \sin^{N-1} \Big(\frac{Jt}{\sqrt{N-1}}\Big), \label{eq:AN1_PST-ex1} \\
   \nonumber \mathcal{A}_{2,N-1} &=& \frac{(-i)^{N-1}}{4} \sin^{N-3} \Big(\frac{Jt}{\sqrt{N-1}}\Big) \times \\
   && \Big(4 - 2 (N - 1) \cos^2 \Big(\frac{Jt}{\sqrt{N-1}}\Big) \Big), \label{eq:AN1_PST-ex2}\\
  \nonumber \mathcal{A}_{3,N-2} &=& \frac{(-i)^{N-1}}{8} \sin^{N-5} \Big(\frac{Jt}{\sqrt{N-1}}\Big) \times \\ && \nonumber  \Big( 8 + \sum_{r = 1}^{2} (-2)^r \cos^{2r} \Big(\frac{Jt}{\sqrt{N-1}}\Big) \times \\
   && \Big( \binom{N-2}{r} + N - 2\Big) \Big),
   \label{eq:AN1_PST-ex3}
\end{eqnarray}
and similarly for the other evolution coefficients. 
The form of the evolution coefficient clearly indicates that for PST, we must have 
\begin{equation}
    J t_{\text{opt}} = (2 n + 1)\sqrt{N - 1} \frac{\pi}{2} ~~~ \text{with} ~ n = 0, 1, \cdots
    \label{eq:optimal_PST_condn}
\end{equation}
At $J t_{\text{opt}}$, each evolution coefficient connecting mirror symmetric modes reduces to $\mathcal{A}_{j, N-j+1} = (-i)^{N-1} \implies |\mathcal{A}_{j, N-j+1}| = 1$ as required for PST (all other elements in each row $j$ of the matrix $\mathcal{A}$ reduce to zero due to normalization). Therefore, the  mode $j$ is transferred to the mode $N - j + 1$, although it acquires a phase of $(-i)^{N-1}$. To complete the PST protocol, we propose the application of a local phase gate, given by $\hat{U}_j = \exp (i \phi_N \hat{a}_j^\dagger \hat{a}_j)$ at each output port $j$. The phase gates at each mode for an $N$-mode waveguide array correspond to the application of the following phases:
\begin{eqnarray}
    \phi_N = \begin{cases}
        \pi ~~ \text{if} ~~ N = 4m - 1, \\
        0 ~~ \text{if} ~~ N = 4m + 1, \\
        \pi/2 ~~ \text{if} ~~ N = 4m + 2, \\
        3\pi/2 ~~ \text{if} ~~ N = 4m. \\
    \end{cases}
    \label{eq:output_phase}
\end{eqnarray}
Therefore, given a $\mathcal{P}$-mode input state in a linear $N$-mode waveguide setup, PST is guaranteed to the mirror-symmetric modes upon evolution for $Jt_{\text{opt}}$ followed by the application of local phase gates at each output mode.

To demonstrate that our protocol indeed realizes PST, let us consider that the input is a single-mode Gaussian state at the first mode and the other $N-1$ modes are initialized as vacuum. \textcolor{black}{The Gaussian state is obtained by applying displacement and squeezing operations to a vacuum state. The displacement introduces a shift in phase space, and squeezing modifies uncertainty in specific quadratures. } The input state is characterized by the displacement vector and covariance matrix, given by
\begin{eqnarray}
    \mathbf{d}_1 = \{\alpha_x, \alpha_y\}, \label{eq:initial_state_d} \\
    \Xi_1  = \begin{pmatrix}
        a & b \\
        b & c
    \end{pmatrix}, \label{eq:initial_state_sigma}
\end{eqnarray}
where $\alpha = \alpha_x + \iota \alpha_y$ is the displacement parameter, and  $a, b, c$ are functions of the squeezing parameter $\tilde{r}$ with $\iota = \sqrt{-1}$. All the parameters are real (for details on the phase-space formalism, see Appendix. \ref{app:CV}. \textcolor{black}{It is essential to highlight that the condition $|\alpha|^2 + e^{-2\tilde{r}} < 1$ must be satisfied for the state to be physically valid, ensuring a well-defined covariance matrix and compliance with the uncertainty principle.} The displacement vector and the covariance matrix pertaining to the  modes $1$ and $N$ during the evolution are given as
\begin{eqnarray}
  &&  \mathbf{d}_1 = \cos^{N - 1}\Big(\frac{J t}{\sqrt{N-1}}\Big) \times \{\alpha_x , \alpha_y \}, \label{eq:d1_any-time}\\
  &&  \Xi_1 = \frac{1}{2} \begin{pmatrix}
        \mathcal{C}_N (a - 1) + 1 & b~\mathcal{C}_N \\
        b~\mathcal{C}_N & \mathcal{C}_N (c - 1) + 1
    \end{pmatrix}, \label{eq:sig1_any-time} \\
     &&  \mathbf{d}_N = \sin^{N - 1}\Big(\frac{J t}{\sqrt{N-1}}\Big) \times \{\alpha_x , \alpha_y \}, \label{eq:dN_any-time} \\
  &&  \Xi_N = \frac{1}{2} \begin{pmatrix}
       \mathcal{S}_N (a - 1) + 1 & b~\mathcal{S}_N \\
        b~\mathcal{S}_N & \mathcal{S}_N (c - 1) + 1
    \end{pmatrix}, \label{eq:sigN_any-time}
\end{eqnarray}
where $\mathcal{C}_N = \cos^{2(N - 1)}\Big(\frac{J t}{\sqrt{N-1}}\Big)$ and $\mathcal{S}_N = \sin^{2(N - 1)}\Big(\frac{J t}{\sqrt{N-1}}\Big)$.
From Eqs. \eqref{eq:d1_any-time} - \eqref{eq:sigN_any-time}, it is evident that at $J t_{\text{opt}}$, $\mathbf{d}_1$ and $\Xi_1$ reduce to those for vacuum whereas the  mode $N$ is characterized by the displacement vector and covariance matrix of the input state, thereby indicating successful implementation of PST. In fact, at the optimal time, the $N$-mode state is described by the following first and second moments, after phase-gate application, as
\begin{eqnarray}
&& \mathbf{d}_f = \{0, 0, \cdots, 0_{2N - 2}, \alpha_x, \alpha_y \} \label{eq:final_d}, \\
  \text{and} && \Xi_{f}=  \frac{1}{2}\Bigg[\mathbb{I}^{\oplus N - 1} \oplus\begin{pmatrix} 
       a & b \\
        b & c
    \end{pmatrix}\Bigg].
    \label{eq:final_cov-mat}
\end{eqnarray}
The input state is taken to be Gaussian and since the evolution is governed by a quadratic Hamiltonian, the evolved state is also so. This allows us to derive the fidelity between the modes $1$ and $N$ during the evolution \cite{Mandarino_IJQI_2014} as
\begin{widetext}
    \begin{eqnarray}
        \mathcal{F} = \frac{2 \exp - \frac{(1 - \mathcal{A}_{N,1})^2 \Big((b \alpha_x^2 - 2 c \alpha_x \alpha_y + a \alpha_y^2)(1 + \mathcal{A}_{N,1}^2) + (\alpha_x^2 + \alpha_y^2)(1 - \mathcal{A}_{N,1}^2)\Big)}{(\chi_+ - 1)(1 + \mathcal{A}_{N,1}^2) + (1 - \mathcal{A}_{N,1}^2)}} {\sqrt{(\chi_-)\mathcal{A}_{N,1}^2 \Big((a + b - 2) + (\chi_+ - a - b)\mathcal{A}_{N,1}^2\Big)} - \sqrt{(\chi_+ + a + b ) + (a + b)\chi_- \mathcal{A}_{N,1}^2 + (\chi_+ - a - b )(\chi_+ - 1)\mathcal{A}_{N,1}^4}}, ~\label{eq:fidelity}
    \end{eqnarray} 
\end{widetext}
where $\chi_\pm = ab - c^2 \pm 1$. Eq. \eqref{eq:fidelity} becomes unity at $J t_{\text{opt}}$. Therefore, with nearest-neighbor couplings defined by Eq. \eqref{eq:opt_J}, any CV Gaussian state can be transferred with perfect fidelity through waveguide arrays comprising an arbitrary number of modes. Looking closely at Eq. \eqref{eq:in_state_fock} and using Eqs. \eqref{eq:AN1_PST-1}, \eqref{eq:optimal_PST_condn} and \eqref{eq:output_phase}, one can show that the same happens for any arbitrary CV state, even if it is non-Gaussian.

\textcolor{black}{\textbf{Remark.} In the discussion above, we considered a single-mode Gaussian input state in a linear waveguide lattice to demonstrate that our formalism can also be explained in the phase-space picture. For such input states, the fidelity function can be explicitly calculated \cite{Mandarino_IJQI_2014}. For  non-Gaussian states, a closed form of the fidelity is not available in the literature, which prevents the analysis of PST for such states intractable in the phase-space formalism. Note, however, that our result extends to any CV state. As demonstrated in Eqs. \eqref{eq:AN1_PST-1} - \eqref{eq:output_phase}, any state comprising an arbitrary number of modes would be perfectly transferred across the lattice at the optimal time. With nearest-neighbor couplings as defined in Eq. \eqref{eq:opt_J}, the condition for PST expressed in terms of the optimal coupling strength, $J t_{\text{opt}}$, ensure that at $t_{\text{opt}}$, each evolution coefficient connecting mirror-symmetric modes reduces to $\mathcal{A}_{j, N-j+1} = (-i)^{N-1} \implies |\mathcal{A}_{j, N-j+1}| = 1$, (and $\mathcal{A}_{j, k} = 0 ~ \forall ~ k \neq N-j+1$ due to the energy-conserving nature of the evolution Hamiltonian), which is the criterion for PST as defined in Eq. \eqref{eq:A_N1_condition}. As our analysis involves the Fock-space description of the input state, it is not restricted to Gaussian states or a single mode.}

\subsection*{Role of entanglement in perfect state transfer} 
It is established that a global entangling operation is necessary for the state transfer protocol. In fact, the evolution through a global Hamiltonian first makes the system (e.g., a lattice chain) genuinely multimode entangled, and then by utilizing that resource, the state is transported from the sender's part to the receiver's end, ultimately destroying the genuine multimode entanglement in the process. However, one-to-one correspondence between the fidelity of state transfer and the entanglement content of the system has not been established, so far.
\begin{figure}
    \includegraphics[width=\linewidth]{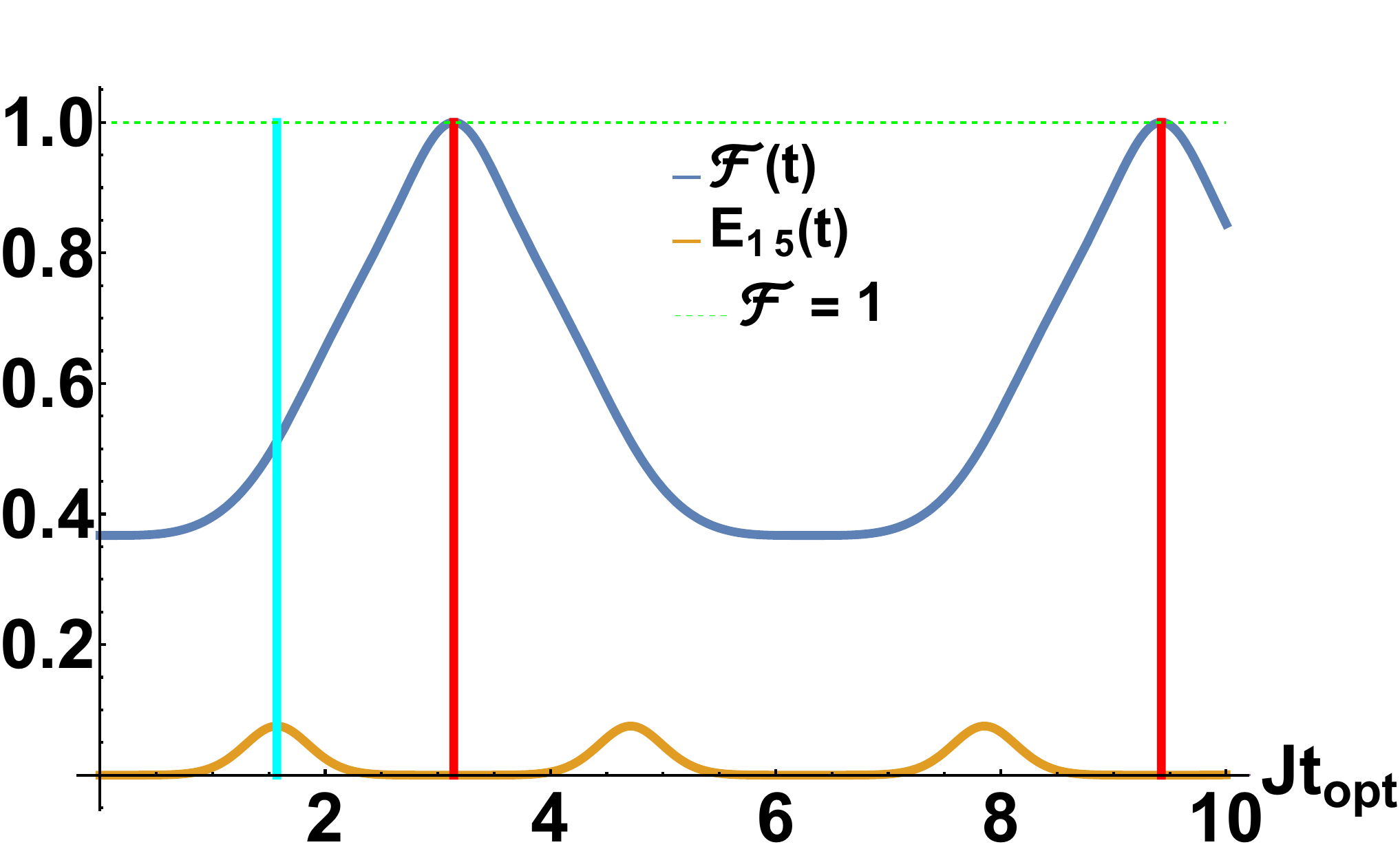}
\captionsetup{justification=Justified,singlelinecheck=false}
     \caption{The behavior of fidelity and entanglement (ordinate) between the input and output modes of a five-mode linear waveguide setup with respect to the evolution parameters $Jt$ (abscissa). The fidelity $\mathcal{F}$ between the states at modes $1$ and $5$, is shown in blue (dark) solid curves, while orange (light) curves represent the logarithmic-negativity quantifying the  entanglement $E_{15}(t)$, between the  modes $1$ and $5$. Green (light) dashed horizontal line shows PST (unit fidelity). The vertical lines identify the values of $Jt = \pi, 3\pi$ (dark red vertical solid line) and $Jt = \pi/2$ (light blue vertical solid line) where the fidelity becomes equal to $1$ and $1/2$ respectively and consequently, the entanglement vanishes and reaches its maximum value. Both axes are dimensionless. }
    \label{fig:fid_ent}
\end{figure}
We analyze the relationship between the dynamics of entanglement shared in the input and output modes and the fidelity. For brevity, we restrict our attention to linear waveguides and the transfer of a single-mode Gaussian state, although qualitatively similar results would hold for multimode CV states, even for non-Gaussian ones. Our treatment focuses on the first and last modes of a linear $N$-mode waveguide arrangement whose evolution is driven by the Hamiltonian in Eq. \eqref{eq:H_NN}. In order to estimate the entanglement, $E_{1N}(t)$ between the input and the output modes during the dynamics, we compute logarithmic negativity \cite{Serafini_2017}, via covariance matrix corresponding to the two-mode state $\rho_{1N}(t) = \Tr_{2,3,\cdots,N-1}\rho(t)$. With the initial state characterized by Eqs. \eqref{eq:initial_state_d} and \eqref{eq:initial_state_sigma} and considering the remaining modes to be initialized in vacuum, the covariance matrix of $\rho_{1N}(t)$ can be represented as 
\begin{eqnarray}
   && \Xi_{1N} = \begin{pmatrix}
        \Xi_1 & \Xi \\
        \Xi & \Xi_N
    \end{pmatrix}, ~~ \text{where} ~~ \\
   && \Xi = \frac{\sin\Big(2 J t/ \sqrt{N-1}\Big)^{N-1}}{2^{N-1}}\begin{pmatrix}
        a-1 & b \\
        b & c-1
    \end{pmatrix},
\end{eqnarray}
and we make use of Eqs. \eqref{eq:d1_any-time} through \eqref{eq:sigN_any-time}. We consider $N = 5$ and the input state parameters as $\alpha_x = \alpha_y = 1, a = 2, b = \sqrt{5}, c = 3$, thereby computing the logarithmic negativity and the fidelity according to Eq. \eqref{eq:fidelity}. We notice that the entanglement between the sender (mode $1$) and the receiver (mode $5$) modes  vanishes when PST occurs at $Jt = Jt_{\text{opt}} = (2n + 1) \pi$, whereas it becomes maximum at  $Jt_{\text{opt}}/2$ as depicted in  Fig. \ref{fig:fid_ent}. This indicates that at the optimal instant, the input and output modes constitute a separable state, a condition that we have argued earlier to be necessary for PST, although entanglement is also a key ingredient that occurs during evolution. Let us also mention here that our results bear a striking resemblance to those in Ref. \cite{Qian_PRA_2005}, obtained for discrete systems, where the entanglement is quantified by the mirror-mode concurrence (which is the sum of the concurrence corresponding to each pair of mirror-symmetric modes) but exhibits the exact same qualitative behavior with respect to the fidelity at $Jt_{\text{opt}}$ and $Jt_{\text{opt}}/2$, even though we consider the logarithmic negativity between the input and the output modes only, regardless of the remaining intermediate modes. Therefore, it can be argued that modal entanglement is certainly responsible for the successful transfer of arbitrary states through the  QST process.

\section{PST leading to mirror-mode swap operation}
\label{sec:sec_swap}

Conventionally, in the state transfer protocol, the modes of the waveguide setup (excluding the input mode) are initially considered to be in the vacuum state. However, even with no input provided, the waveguide modes are not expected to be in the vacuum state and as such need to be engineered to $\ket{0}$. Since the Hamiltonian in Eq. \eqref{eq:H_NN} together with appropriate phase gates can perfectly transfer states between modes $j$ and $N-j+1$, we can also employ our protocol to realize SWAP gates between the aforementioned modes. In this case, it is evident that the user must have access to all the waveguide modes since the states to be interchanged must be provided apriori to the evolution.

Let us consider that the modes $2$ to $N$ in the network are initialized with any arbitrary Gaussian state which can be characterized by the displacement vector and covariance matrix as

\begin{eqnarray}
  \nonumber &&  \mathbf{d}_{\text{in}} = \{\alpha_x, \alpha_y\}^T \oplus_{i = 2}^N \{\beta_{xi}, \beta_{yi}\}^T, \\
  \text{and} && \Xi_{\text{in}} = \begin{pmatrix}
        a & b \\
        b & c
    \end{pmatrix} 
    \oplus_{i = 2}^N \begin{pmatrix}
        p_i & q_i \\
        q_i & s_i
    \end{pmatrix},
    \label{eq:arb_state_in-phase-space}
    \end{eqnarray}
where $\{\alpha_x, \alpha_y\}^T$ corresponds to the displacement vector of the input mode and $\{\beta_{xi}, \beta_{yi}\}^T$ denotes the displacement vector of the arbitrary state in mode $i$. Similarly, $a, b,$ and $c$ are the elements of the covariance matrix pertaining to the input state, whereas $p_i, q_i$ and $s_i$ stand for the same in the case of the state in mode $i$. Now, upon evolution through an $N$-mode waveguide characterized by a Hamiltonian with nearest-neighbor couplings defined in Eq. \eqref{eq:opt_J}, the final state at  $J t_{\text{opt}} = \sqrt{N - 1}\frac{\pi}{2}$ followed by the application of corresponding phase gates, as prescribed in Eq. \eqref{eq:output_phase}, in all the modes,  is described by
\begin{eqnarray}
  \nonumber &&  \mathbf{d}_f = \oplus_{i = 1}^{N-1}  \times \{\beta_{xi}, \beta_{yi}\}^T \oplus \{\alpha_x, \alpha_y\}^T, \\
   && \Xi_f = \oplus_{i = 1}^{N-1} \begin{pmatrix}
        p_{N-i+1} & q_{N-i+1} \\
        q_{N-i+1} & s_{N-i+1}
    \end{pmatrix} 
     \oplus \begin{pmatrix}
       a & b \\
       b & c
    \end{pmatrix},
    \label{eq:arb_state_out-phase-space}
   \end{eqnarray}
 which clearly illustrates 
that $\Xi_f$ is just a mirror-inverted version of $\Xi_{\text{in}}$ about the central mode when the total number of modes is odd ($N = 2m + 1$) or about the central coupling when the total number of modes is even ($N = 2m$). Note that the protocol still goes through for any arbitrary initialization and Gaussian states are taken to illustrate the fact in the phase space formalism. Thus, by applying local phase shifts in all the modes, one can realize the SWAP operations between mirror-symmetric modes by evolving the system for a period of $(0, t_{\text{opt}})$. Therefore, if required states are provided in specific modes, a collateral benefit of the optimal PST Hamiltonian is to act as a SWAP operator between modes that are equispaced from the center of the waveguide array. This, however, incurs the cost of local phase shifters in each mode, unlike the PST process where only the output mode(s) need to be phase-shifted. 

\textbf{Remark: }Let us consider that the initial state of an $N$-mode system is $\oplus_{j = 1}^{N} |\psi^{j}\rangle_{j} = \oplus_{j = 1}^{N} \Big(\sum_{n_k = 0}^{\infty} C^{j}_{n_{k}} |n_k\rangle_j\Big)$, where the superscript $j$ represents the state label, while $j$ in the subscript denotes the mode at which it is present, and $\ket{n_k}$ is the state containing $n_k$ number of photons. Upon evolution up to the optimal time and further implementation of local phase-shifts, the final state reduces to $\oplus_{j = 1}^{N} \Big(\sum_{n_k = 0}^{\infty} C^{j}_{n_{k}} |n_{k}\rangle_{N-j+1}\Big) = \oplus_{j = 1}^{N} |\psi^j\rangle_{N-j+1}$, which is a consequence of $|\mathcal{A}_{j,N-j+1}| = 1~~\forall~~j$ (a condition that depends on the Hamiltonian in action and not on the initialization of the lattice). Therefore, mirror-mode SWAP operation can be implemented for a collection of arbitrary CV states.

\section{PST in higher dimensional lattices}
\label{sec:2d-PST}

Going beyond one-dimensional systems, we demonstrate how PST of any arbitrary CV state can be implemented on two-dimensional ($2$D) and three-dimensional ($3$D) waveguide lattices. We use the PST Hamiltonian in \eqref{eq:H_NN} to derive the evolution which would lead to PST from one part to another of a waveguide setup arranged in $2D$ and $3D$. Accurate transfer of a CV state in a higher dimensional lattice would allow us to realize state transfer across several modes in a more compact manner since distant transfer along a linear array would require a very long chain of waveguide modes.

\begin{figure}
    \includegraphics[width=\linewidth]{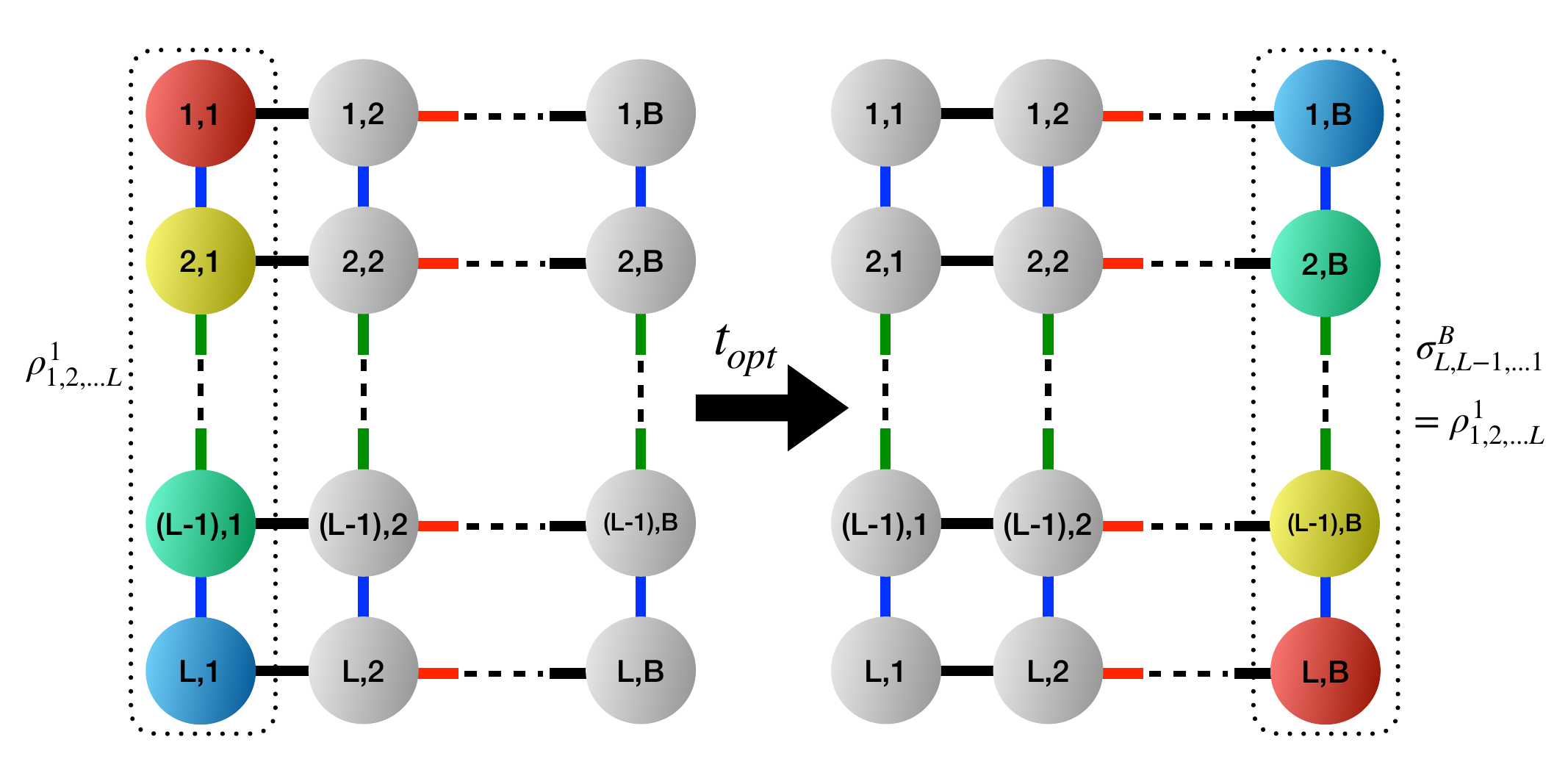}
\captionsetup{justification=Justified,singlelinecheck=false}
     \caption{\textbf{Two-dimensional lattice of waveguide modes supporting PST}. The arrangement consists of $\mathcal{L}$ rows and $\mathcal{B}$ columns, with modulated nearest-neighbor couplings which are mirror symmetric. Each available mode is denoted as $i,j$. The input state, $\rho^1_{1,2,...,L}$ at the modes $1,1$ to $1, L$ can be a separable or entangled $L$-mode state, and the remaining modes are initialized with the vacuum states, but can be taken to be any arbitrary state. Suitable local phase gates, following the evolution under the specified Hamiltonian for time $t_{opt}$, must be implemented at the receivers' end (at the far right side) in order to accomplish PST - when the output state $\sigma^\mathcal{B}_{1,2, \cdots, L}$ at the modes $1,\mathcal{B}$ to $\mathcal{L}, \mathcal{B}$ (which are mirror-symmetrically opposite to the input modes about the center of the geometry) becomes exactly the same as the input one, thereby leading to PST. }
    \label{fig:sch_pst_2D}
\end{figure}

We define our coordinate system as follows. $\mathcal{L}$ and $\mathcal{B}$ modes along two orthogonal directions comprise a $2D$ latttice on a plane. Along the direction perpendicular to the plane, we consider $\mathcal{H}$ number of waveguide modes when in $3D$. 

\subsubsection{PST in two-dimensional waveguide lattice}
\label{subsubsec:2d}
Let us first concentrate on a rectangular cluster with $\mathcal{L}$ rows and $\mathcal{B}$ columns, as illustrated in Fig. \ref{fig:sch_pst_2D}. Therefore, the total number of modes supported by the entire structure is $N = \mathcal{L} \times \mathcal{B}$. Evidently, we can consider each row as an $\mathcal{B}$-mode linear waveguide chain, which is coupled to another linear pattern of $\mathcal{B}$ waveguide modes. Conversely, an $\mathcal{L}$-mode linear waveguide setup describes every column of the rectangular system. Every mode within the lattice is linked with four other modes, except for the modes at the boundaries and corners, which have a coordination number of $3$ and $2$ respectively.
The Hamiltonian governing the PST protocol is $\hat{H}^{\mathcal{L},\mathcal{B}} = \hat{H}^{\mathcal{L}} + \hat{H}^{\mathcal{B}}$ where

\begin{eqnarray}
   && \hat{H}^{\mathcal{L}} = \sqrt{\frac{\mathcal{L}-1}{\mathcal{B} - 1}} \sum_{j,k = 1}^{\mathcal{L} -1, \mathcal{B}} J^{\mathcal{L}}_j (\hat{a}_{(j,k)}^\dagger \hat{a}_{(j+1,k)} + \hat{a}_{(j,k)} \hat{a}_{(j+1,k)}^\dagger), ~~~~~~ \label{eq:ham_2d_column}  \\
  && \text{and} ~~ \hat{H}^{\mathcal{B}} = \sum_{j,k = 1}^{\mathcal{L}, \mathcal{B} - 1} J^{\mathcal{B}}_k (\hat{a}_{(j,k)}^\dagger \hat{a}_{(j,k+1)}  + \hat{a}_{(j,k)} \hat{a}_{(j,k+1)}^\dagger). \label{eq:ham_2d_row}
\end{eqnarray}
Evidently, the Hamiltonian to realize PST in a $2D$ lattice comprises a  Hamiltonian for $\mathcal{L}$ number of modes, whose interaction strength is scaled by a factor of $\sqrt{\frac{\mathcal{L}-1}{\mathcal{B} - 1}}$, and another one-dimensional Hamiltonian corresponding to $\mathcal{B}$ modes. \textcolor{black}{The Hamiltonians along different axes commute, i.e., $[\hat{H}^\mathcal{L}, \hat{H}^\mathcal{B}] = 0$. Thus, for $\hat{H} = \sum_{j = \mathcal{B}, \mathcal{H}} \hat{H}^j$, we can write the evolution operator as $e^{i \hat{H} t} = \prod_{j = \mathcal{B}, \mathcal{L}} e^{i \hat{H}^j t}$, which implies that the time evolution occurs independently across the two directions. The scaling factor, $\sqrt{\frac{\mathcal{L}-1}{\mathcal{B} - 1}}$, for the coupling strength along the $\mathcal{L}$ direction ensures that the information propagates equal distance in each direction for a given time interval, leading to PST between mirror-symmetric modes. A similar argument also applies to three-dimensional lattices with NN couplings, which will be discussed later.} Setting $J^{\mathcal{B}}_k$ and $J^{\mathcal{L}}_j$ satisfying Eq. \eqref{eq:opt_J}, 
one can realize PST from the mode $(i,j)$ to the mode $(\mathcal{L}-i+1, \mathcal{B}-j+1)$ at the mirror-symmetric position of the waveguide composition aided with local phase gates at the output modes.

The evolution coefficients, involving mirror-symmetric sites, and the optimal time bear resemblance to the ones for the linear chain, as 

\begin{eqnarray}
   && \mathcal{A}_{(j,k),(\mathcal{L}-j+1, \mathcal{B}-k+1)} = \mathcal{A}_{j,\mathcal{L} - j+1} \times \mathcal{A}_{k,\mathcal{B} - k+1}  \label{eq:An1-2d}\\
   &&  \text{with} ~~  J t_{\text{opt}} = (2 n + 1) \sqrt{\mathcal{B} - 1} \frac{\pi}{2} ~~ (n = 0, 1, \cdots,). ~~~~~~ \label{eq:opt-t_2d}
\end{eqnarray}
The fact that the evolution coefficient in the case of a $2D$ waveguide system can be factored into a product of the same in each involved direction follows from the property that the state transfer proceeds independently in separate dimensions \cite{Kay_IJQI_2010}. Note that the evolution coefficients must be evaluated according to Eqs. \eqref{eq:ham_2d_column} and \eqref{eq:ham_2d_row}. This would lead to expressions for $\mathcal{A}_{j,\mathcal{L} - j+1}$ and $\mathcal{A}_{k,\mathcal{B} - k+1}$ with functional forms similar to Eq. \eqref{eq:AN1_PST-1} but with arguments scaled by the aforementioned factor, if any. Moreover, the phase acquired by each output mode during evolution is $(-i)^{\mathcal{L} + \mathcal{B} - 2}$ in the case of a $2D$ waveguide lattice. Therefore, the phase gates to be provided at the output modes correspond to implementing a correction phase given by
\begin{eqnarray}
    \phi_{\mathcal{L}\mathcal{B}} = \begin{cases}
        \pi/2 ~~ \text{if} ~~ \mathcal{L} + \mathcal{B} = 4m - 1 \\
        3 \pi/2 ~~ \text{if} ~~ \mathcal{L} + \mathcal{B} = 4m + 1 \\
        0 ~~ \text{if} ~~ \mathcal{L} + \mathcal{B} = 4m + 2 \\
        \pi ~~ \text{if} ~~ \mathcal{L} + \mathcal{B} = 4m. \\
    \end{cases}
    \label{eq:output_phase-2d}
\end{eqnarray}

It is thus clear that PST occurs from the site $(j,k)$ to the mode at site $(\mathcal{L}-j+1, \mathcal{B}-k+1)$ when the evolution is executed for an optimal time given in Eq. \eqref{eq:opt-t_2d} followed by a phase shift at the output modes. Therefore, applying  the PST Hamiltonian for linear waveguide setups, one can also achieve the transfer of any arbitrary state, across a $2D$ rectangular lattice, with unit fidelity, notwithstanding the initialization of the non-input modes in the waveguide. If a multimode (say, $\mathcal{P}$ mode) CV state, be it pure or mixed, Gaussian or non-Gaussian, is to be transmited, it would be transferred perfectly to the $\mathcal{P}$ mirror-symmetric modes at the output. With all the $\mathcal{P}$ output modes being accessible, the application of phase gates, which is essential to achieve unit fidelity, can also be easily applied. Furthermore, $\hat{H}^{\mathcal{L} \mathcal{B}}$ also acts as a SWAP gate between modes $(j,k)$ and $(\mathcal{L} - j + 1, \mathcal{B} - k + 1)$ when local phase gates are applied in all the modes in the lattice as discussed in the previous section. As a result, the Hamiltonian governing PST in a $2D$ arrangement of waveguide modes retains all the beneficial properties of that concerning linear PST.

\subsubsection{Three-dimensional waveguide arrangement for PST}
\label{subsubsec:3d}
The Hamiltonian which supports the perfect transfer of any arbitrary multimode CV state across a $3D$ geometry can be written as $\hat{H}^{\mathcal{L}\mathcal{B}\mathcal{H}} = \hat{H}^{\mathcal{L}} + \hat{H}^{\mathcal{B}} + \hat{H}^{\mathcal{H}}$ with
\begin{equation}
    \hat{H}^{\mathcal{H}} = \sqrt{\frac{\mathcal{H}-1}{\mathcal{B} - 1}} \sum_{j,k,l = 1}^{\mathcal{L} , \mathcal{B}, \mathcal{H}-1} J^{\mathcal{H}}_l (\hat{a}_{(j,k,l)}^\dagger \hat{a}_{(j,k,l+1)} + \hat{a}_{(j,k,l)} \hat{a}_{(j,k,l+1)}^\dagger), ~~~~~~ \label{eq:ham_3d_height} 
\end{equation}
and $\hat{H}^{\mathcal{L}}$ and $\hat{H}^{\mathcal{B}}$ given by Eqs. \eqref{eq:ham_2d_column} and \eqref{eq:ham_2d_row} where the summations run over the three indices $j,k,l$ along the three-dimensions. \textcolor{black}{The existence of the scaling factor in $\hat{H}^{\mathcal{H}}$ is due to the reasons mentioned previously for the two-dimensional lattice.} The waveguide arrangement contains a total number of modes equal to $N = \mathcal{L}\mathcal{B}\mathcal{H}$. Akin to the two-dimensional QST protocol, the evolution of the system is independent in each dimension and we can once more write down the evolution coefficient as a product of the same in each direction. Evaluating the evolution coefficients corresponding to $\hat{H}^{\mathcal{L}}, \hat{H}^{\mathcal{B}},$ and $\hat{H}^{\mathcal{H}}$, it is easy to see that in $3D$, the optimal evolution coefficient is $\mathcal{A}_{(j,k,l),(\mathcal{L}-j+1, \mathcal{B}-k+1, \mathcal{H} - l +1}) = \mathcal{A}_{j,\mathcal{L} - j+1} \times \mathcal{A}_{k,\mathcal{B} - k+1} \times \mathcal{A}_{l,\mathcal{H} - l+1}  ~~ \text{with} ~~  J t_{\text{opt}} = (2 n + 1) \sqrt{\mathcal{B} - 1} \frac{\pi}{2} ~~ \text{for}~~ (n = 0, 1, \cdots,)$. It is observed that evolving the initial system for the optimal time reduces each $\mathcal{A}_{(j,k,l),(\mathcal{L}-j+1, \mathcal{B}-k+1, \mathcal{H} - l +1)} ~~ \text{to} ~~ (-i)^{\mathcal{L} + \mathcal{B} + \mathcal{H} - 3}$ which represents the local phase acquired by each mode. This calls for the application of phase gates at the output sites corresponding to
\begin{eqnarray}
    \phi_{\mathcal{L}\mathcal{B}\mathcal{H}} = \begin{cases}
        \pi/2 ~~ \text{if} ~~ \mathcal{L} + \mathcal{B} + \mathcal{H} = 4m,\\
        3 \pi/2 ~~ \text{if} ~~ \mathcal{L} + \mathcal{B} + \mathcal{H} = 4m + 2,\\
        0 ~~ \text{if} ~~ \mathcal{L} + \mathcal{B} + \mathcal{H} = 4m - 1, \\
        \pi ~~ \text{if} ~~ \mathcal{L} + \mathcal{B} + \mathcal{H} = 4m + 1, \\
    \end{cases}
    \label{eq:output_phase_3d}
\end{eqnarray}
as the correction phase, in order to ensure state transfer to the mirror-symmetric output modes with unit fidelity. Similar to the cases in $1D$ and $2D$, the application of proper local phase shifts in all the waveguide modes can help to use the Hamiltonian in implementing a SWAP gate between modes $(j,k,l)$ and $(\mathcal{L} - j + 1, \mathcal{B} - k + 1, \mathcal{H} - l + 1)$. Therefore, if the condition for PST is satisfied by $\hat{H}^{\mathcal{L} \mathcal{B} \mathcal{H}}$, evolution through $t = 0 \rightarrow t_{\text{opt}}$ followed by local phase gates can ensure the realization of PST as well as mirror-mode SWAP operation of any multimode CV state, irrespective of its Gaussian (non-Gaussian) character and purity. 

\textcolor{black}{\textbf{Remark.} The transfer matrix $\mathcal{A}$ is given by $\mathcal{A} = \exp (-i \mathcal{M} t)$, where $\mathcal{M}$ is the adjacency matrix corresponding to the evolution $\hat{H}$. Consequently, the analytical expression for $\mathcal{A}$ depends on the Hamiltonian $\hat{H}$. Note that, in our work, we are only concerned with PST between mirror symmetric modes and are thus not concerned about the other coefficients of the transfer matrix. For a Hamiltonian with nearest-neighbor couplings specified by Eq. \eqref{eq:opt_J}, the transfer matrix elements corresponding to mirror-symmetric modes have been specified for a one-dimensional lattice in Eq. \eqref{eq:AN1_PST-1}. For higher dimensional lattices, one can easily derive the same, since in such cases, the mirror-symmetric coefficients of $\mathcal{A}$ are a product of the same in each considered dimension.}

\textcolor{black}{\textbf{Note.} The phase space analysis similar to the PST of a single-mode Gaussian state in a linear array can again be repeated for higher dimensional lattices with some more complicated algebra. However, the fidelity expression for arbitrary multimode Gaussian states is analytically intractable which restricts us from examining PST of multimode Gaussian states in the phase space formalism. However, for any input state (Gaussian or non-Gaussian), we can resort to analysis in the Fock basis which is explicitly conducted in our work.}

\subsection*{Possible experimental implementation of PST Hamiltonian}
\label{subsec:expt}

Optical lattices serve as a potential framework for realizing tight-binding Hamiltonians \cite{Christodoulides_Nature_2003, Longhi_LPR_2009, Szameit_JPB_2010} for large-scale quantum information processing \cite{Politi_Science_2008}. Thus, our proposed Hamiltonian, belonging to the tight-binding class can be implemented in waveguide arrays which, moreover, admit negligible photon interactions, provide long coherence times, and also allow for high experimental control \cite{Owens_NJP_2011}. A mode-dependent coupling strength, like the one we present for our purpose, can be engineered by suitably manipulating the waveguide separation \cite{Szameit_JPB_2010}. In fact, for each $J_{j}$ as defined in Eq. \eqref{eq:opt_J}, (with additional factors in $2D$ and $3D$), the optimal separation $\kappa_j$ between waveguide modes $j$ and $j + 1$, in the weak coupling limit, can be doctored according to
\begin{equation}
    \kappa_j = \eta^{-1} \Big( \ln \frac{\gamma}{J} - \ln \frac{J_{j}}{J} \Big),
    \label{eq:expt_J-opt}
\end{equation}
where $\gamma$, and $\eta$ are parameters depending on the particular setup \cite{Bellec_OL_2012} and $J$ as defined in Eq. \eqref{eq:opt_J}. Note that the above formalism is applicable only when the coupling strengths between the adjacent modes are small, which is ensured since each $J_j$ is scaled by $1/\sqrt{N-1}$, i.e., for large $N$, even though the numerator in Eq. \eqref{eq:opt_J} increases, so does the scaling factor and the coupling strengths remain experimentally manageable. It is interesting to mention here that diffraction losses occurring in any waveguide array due to lattice truncation and edge effects \cite{Longhi_PRB_2010} can be readily minimized by engineering the coupling coefficients \cite{Gordon_OL_2004, Keil_OL_2012} which is the essence of our PST protocol. As shown in reference \cite{Fukuda_ISOP_2004}, negligible propagation defects make waveguide structures a feasible breadboard for implementing the PST Hamiltonian effectively.

\section{Conclusion}
\label{sec:conclu}
In a quantum computing circuit, state transfer can be a crucial component for propagating quantum information from one quantum processor to another. The information transmitted should also be as little distorted as possible, to ensure smooth functioning of the network. On the other hand, quantum state transmission is also an integral part of quantum communication networks. Therefore, finding an optimal protocol for perfect quantum transfer (PST) is one of the demanding issues in both the domains of quantum computers and quantum communication.

 We demonstrated the perfect transfer of any given continuous variable (CV) quantum state through an optical waveguide lattice governed by a  Hamiltonian comprising modulated nearest-neighbor couplings. In particular, we found a criterion that should be met by any Hamiltonian required during evolution in order to accurately transfer an arbitrary state from one part to another in an optical waveguide array. \textcolor{black}{There are several contributions in the literature on PST in  CV systems, e.g., the perfect transfer of single- and two-photon Fock states, or pure path entangled states.} The perfect transfer of non-classical properties such as squeezing strength as well as average photon number has also been reported \cite{Rai_JO_2022}. It is important to note, however, that transmitting photon numbers accurately does not translate into PST, since two states differing by local phases can have the same mean number of photons. This is significant since such states are essentially different from each other since they can carry vastly different information.
 
 \textcolor{black}{In this work, we studied the PST of a generic multimode CV state, which may be pure or mixed implying irrevocably precise transfer of the encoded information.} Our proposed protocol, with mirror-symmetric Hamiltonian and the subsequent application of specific phase-shifts, can further allow for the relocation of any CV state between the two mirror-symmetric of a waveguide setup in two- as well as three-dimensions, with unit fidelity, upon evolution up to some optimal time. \textcolor{black}{The coupling strengths are derived according to the approach outlined in Ref. \cite{Christandl_PRL_2004} which supports PST in a spin chain, and thus, an equivalent lattice of coupled waveguides with no nonlinearity may similarly support PST of any state of the electromagnetic field. Our protocol demands the application of local phase shifts at the output modes, the correction phase being independent of the input state to be transferred and only depending on the total number of modes comprising the waveguide arrangement. We argued that this local manipulation is allowed since the input and output modes are always accessible in a PST experiment.  In addition, we showed that with the help of local phase unitaries in all the waveguide modes, a mirror-mode swapping of states can be achieved as a collateral benefit of our engineering. One of the main features of our work is that we do not need to bother about the initial state of modes other than the input ones. Indeed, any initialization of the system (other than the sender mode) can accomplish PST.} We further looked into how the entanglement between the sender and the receiver modes of the linear waveguide array is related to the efficiency of the protocol. More precisely, we noted that the aforementioned entanglement is maximum at half the time taken to achieve  PST, whereas it vanishes at the instant of PST. This showed how entanglement is an indispensable component for the success of the protocol, in which the assistance of entanglement is still present.

Our work thus extends the theory of continuous-variable state transfer to bring all continuous-variable states under one umbrella. Our protocol is experimentally feasible and would lead to the successful implementation of several information theoretic protocols driven by linear optical elements \cite{OBrien_Science_2007}, which rely on state transfer. 

\section*{Acknowledgement}
\label{sec:acknowledgement}
We acknowledge the support from the Interdisciplinary Cyber-Physical Systems (ICPS) program of the Department of Science and Technology (DST), India, Grant No.: DST/ICPS/QuST/Theme- $1/2019/23$. This research was supported in part by the 'INFOSYS scholarship for senior students'. R.G. acknowledges funding from the HORIZON-EIC-$2022$-PATHFINDERCHALLENGES-$01$ program under Grant Agreement No.~$10111489$ (Veriqub). Views and opinions expressed are however those of the authors only and do not necessarily reflect those of the European Union. Neither the European Union nor the granting authority can be held responsible for them.

\appendix
\section{CV-systems - a brief discussion on the phase-space formalism}
\label{app:CV}
Characterized by quadrature variables, such as $\hat{x}$ and $\hat{p}$, which are canonically conjugate with each other, continuous variable systems  possess an infinite spectrum \cite{Serafini_2017, Braunstein_RMP_2005}. The Hamiltonian of an $N$-mode system comprising $2N$ parameters, $\{\hat{x}_k, \hat{p}_k\}$ (with $k = 1,2,\dots, N$), is defined as
\begin{equation}
    \hat{H} = \frac{1}{2} \sum_{k = 1}^N (\hat{x}_k^2 + \hat{p}_k^2) = \sum_{k = 1}^N \Big(\hat{a}_k^\dagger \hat{a}_k + \frac{1}{2} \Big),
    \label{eq:CV_hamiltonian}
\end{equation}
where $\hat{a}_k^\dagger$ and $\hat{a}_k$ are the creation and annihilation operators respectively for the mode $k$ and, in terms of the quadrature variables, they are given by
\begin{equation}
    \hat{a}_k = \frac{\hat{x}_k + i \hat{p}_k}{\sqrt{2}}, ~~~~~~~ \text{and} ~~~~~~ \hat{a}_k^\dagger = \frac{\hat{x}_k - i \hat{p}_k}{\sqrt{2}},
    \label{eq:creation-annihilation_op}
\end{equation}
with $i = \sqrt{-1}$. The bosonic commutation relation, $[\hat{a}_k^\dagger, \hat{a}_k] = -1$, is obeyed by the creation and annihilation operators for any mode. The quadrature vector, $\hat{R} = (\hat{x}_1, \hat{p}_1, \dots, \hat{x}_N, \hat{p}_N)^T$, helps to rewrite the commutation relation as
\begin{equation}
    \left[\hat{R}_k,\hat{R}_l\right]=i \mathcal{W}_{kl}\quad \text{with} ~~  \mathcal{W} = \bigoplus\limits_{j=1}^{\mathcal{N}} \Omega_j.
  \label{eq:CV_commutation}
  \end{equation}
 Here, $\mathcal{W}$ represents the $\mathcal{N}$-mode symplectic form, and $\Omega_j$ is given by
  \begin{equation}
      \quad \Omega_j=\begin{pmatrix}
			0 & 1\\
			-1 & 0 
		\end{pmatrix} \forall j.
  \label{eq:CV_omega}
	\end{equation}
 Gaussian states are the most widely studied class of CV states \cite{ferraro2005, Weedbrook_RMP_2012}. Being the ground and thermal states of Hamiltonians which are at most quadratic functions of the quadrature variables, a Gaussian state, $\rho$, can be completely characterized by their first and second moments, quantified respectively by the $2N$-element displacement vector $\mathbf{d}$ and the $2N \times 2N$ dimensional covariance matrix, $\Xi$, in the following way:
 \begin{eqnarray}
		&& d_k=\expval{\hat{R}_k}_{\rho}, \\
  \label{eq:CV_disp}
		\Xi_{kl}&&= \frac{1}{2}\expval{\hat{R}_k\hat{R}_l+\hat{R}_l\hat{R}_k}_{\rho}-\expval{\hat{R}_k}_{\rho}\expval{\hat{R}_l}_{\rho}.
  \label{eq:CV_cov}
	\end{eqnarray}
 Here, $\Xi$ is a real, symmetric, and positive definite matrix. Gaussian dynamics is governed similarly by second-order Hamiltonians. In the symplectic formalism, any $N$-mode quadratic Hamiltonian $\hat{\mathcal{H}}$ can be written as $\hat{\mathcal{H}}=\frac{1}{2} \hat{\xi}^\dagger \hat{H} \hat{\xi}$ with $\hat{\xi}=\left(\hat{a}_1, \hat{a}_2,..., \hat{a}_N, \hat{a}_1^\dagger,..., \hat{a}_N^\dagger\right)^T$. We can construct the symplectic matrix $S_H$ as \cite{Luis_QSO_1995, Arvind_Pramana_1995, Adesso_OSID_2014}
 \begin{eqnarray}
     S_H = T^\dagger L^\dagger \exp{- i K \hat{H} } L T,
     \label{eq:H_symp}
 \end{eqnarray}
 where $K, L$ and $T$ are $2N \times 2N$ matrices given by 
 \begin{eqnarray}
    && K = \begin{pmatrix}
         \mathbb{I}_N & \mathbb{O}_N \\
         \mathbb{O}_N & -\mathbb{I}_N
     \end{pmatrix}, \label{eq:K_mat} \\
    && L = \frac{1}{\sqrt{2}}\begin{pmatrix}
         \mathbb{I}_N & i \mathbb{I}_N \\
         \mathbb{I}_N & -i \mathbb{I}_N
     \end{pmatrix}, \label{eq:L_mat} \\
    && T_{jk} = \delta_{k,2j - 1} + \delta_{k + 2N, 2j} \label{eq:T_mat},
 \end{eqnarray}
 with $\mathbb{I}_N$ being the $N$-dimensional identity and $\mathbb{O}_N$ being the null matrix.
 The evolved Gaussian state, in terms of its displacement vector and covariance matrix, can be characterized as \cite{Adesso_OSID_2014}
 \begin{eqnarray}
 \rho' =  e^{-i \hat{H}t} \rho e^{i \hat{H} t} \equiv && \bold{d}' = S_H \bold{d}, \label{eq:evolved_disp} \\
 && \Xi' = S_H \Xi S_H^T, \label{eq:evolved_cov}
 \end{eqnarray}
where the evolution is governed by $\hat{H}$ in Eq. \eqref{eq:CV_hamiltonian} and since the evolved state also remains Gaussian, it can be represented by displacement and covariance matrix of the initial state via $S_{H}$ in Eq. \eqref{eq:H_symp}

 \bibliographystyle{apsrev4-1}
\bibliography{bib}

\end{document}